# Topological Complexity of Frictional Interfaces: Friction Networks

H. O. Ghaffari, R. P. Young

*Department of Civil Engineering and Lassonde Institute,*
*University of Toronto, Toronto, ON, Canada*

Through research conducted in this study, a network approach to the correlation patterns of void spaces in rough fractures (crack type II) was developed. We characterized friction networks with several networks characteristics. The correlation among network properties with the fracture permeability is the result of friction networks. The revealed hubs in the complex aperture networks confirmed the importance of highly correlated groups to conduct the highlighted features of the dynamical aperture field. We found that there is a universal power law between the nodes' degree and motifs frequency (for triangles it reads $T(k) \propto k^{\beta}$ ($\beta \approx 2 \pm 0.3$)). The investigation of localization effects on eigenvectors shows a remarkable difference in parallel and perpendicular aperture patches. Furthermore, we estimate the rate of stored energy in asperities so that we found that the rate of radiated energy is higher in parallel friction networks than it is in transverse directions. The final part of our research highlights 4 point sub-graph distribution and its correlation with fluid flow. For shear rupture, we observed a similar trend in sub-graph distribution, resulting from parallel and transversal aperture profiles (a superfamily phenomenon).

## Table of Contents

# Introduction

Cracks, fractures, joints and – on a global scale – faults are the main elements of fluid flow and induced disorder in rock masses. The evolution of such frictional interfaces and the onset of slip are integrated with the developments of the contact areas (Dieterich and Kilgore, 1994; Rubinstein et al, 2004 and 2009; Thompson et al, 2009). The crack-like behaviour of rupture in frictional interfaces also supports the role of relative contact areas and apertures (Rubinstein et al, 2004 and 2009; Ben-David and Fineberg, 2010). Also, the variations of fluid flow features (such as permeability and tortuosity) are controlled directly with aperture spaces (Auradou et al, 2005; Sharifzadeh, 2005; Ghaffari et al, 2011b). Characterization of contact patterns and possible correlation among elements of the sheared system are the key elements in the analysis of sheared systems as well as frictional interfaces. Such characterization has been realized by using several techniques such as employing statistical methods that include root-mean square (RMS), RMS of first derivative (Z2), RMS of second derivative (Z3), and structure function (SF) (Fardin et al ,2001; Lanaro and Stephansson ,2003 ;Sharifzadeh ,2005) ; geo-statistical methods to study spatial variation of asperity heights as well as spatial correlation (semi-variograms) and ; correlation length which gives an idea of the aperture changes over the frictional surface; the fractal models mainly used for scale effect analysis (Lanaro and Stephansson ,2003). We notice nonlinear and collective behaviours of contact areas in frictional interfaces indicate a complex system with intricate response to environment stimuli.

One of the recent theories to analysis of complex systems is network theory. Network theory is a fundamental tool for the modern understanding of complex systems in which, by a simple graph representation, the elementary units of a system become nodes, and their mutual interactions become links. With this transformation of a system to a network space, many properties about the structure and dynamics of the system itself can be inferred. Recently, direct network approaches (from mechanical point of view) have been used to study the behaviour of crumpled papers, shells (Aharoni and Sharon,2010; Andresen et al, 2007) and force chains in granular materials (Tordesillas et al, 2010) .The indirect approaches mostly transfer the information in time series (recurrent events) into graphs, and analysis obtained graphs with networks attributes (for example see: Donges et al ,2009 ; Gao et al ,2010). Earthquake networks resulting from faults activities, which are another form of indirect networks, have also been addressed (Abe and Suzuki, 2006; Baiesi and Paczuski, 2004). We notice earthquake are the direct results of evolution of frictional interfaces and then analysis emerged patterns of earthquakes give insight into the fault mechanism .In this approach, the events are connected to each other based on a novel metric that includes Euclidean distance of events and the spatial heterogeneity. Analysis of earthquake networks showed that networks could be useful in detecting symptoms and signatures that typically precede events. Evolution of earthquake network parameters also gives rise to a close-up point of view to underlying earthquake dynamics, i.e., characterizing complex phenomena such as earthquakes. With respect to single fracture behaviour, the opening spaces (i.e., aperture patches) were mapped onto networks based on a Euclidean metric (Ghaffari et al, 2009 and 2010). The results showed the clustering coefficient of obtained networks roughly to scale with the mechanical or hydraulic properties of the shear fracture.

In the present study, we mapped the measured patterns of aperture patches onto the network, using an indirect network approach. With this transformation of aperture patches in networks, we introduced friction networks. The complex contact area patterns result from the careful measurement of the rough fracture surfaces with a scanner laser over different cases of normal loads. Our focus was on the possible correlation among networks' characteristics and the mechanical and hydro-mechanical features (i.e., permeability) of the fracture. Remarkably, we found a universal power law among the nodes' degree and the frequency of loops. Based on this finding, we extended the classical rate and state friction law in terms of network evolution. Next, we investigated rate of energy storage in friction networks. To do this, we distinguished the synchronization of energy flux through the obtained networks, noting its relation to the characteristic length of networks as well as to the localization of Laplacian matrix eigenvectors. This transformation of correlation patterns to network spaces can be interpreted as a way to take into account the long-range interactions of profiles, as well as the force chains in granular materials. A further part of our research was the motif analysis of the constructed networks. In our study, analysis of 4-node sub-graphs revealed a superfamily phenomenon in sheared interfaces. In other words, a similar trend in sub-graphs distribution resulted from parallel and transversal aperture profiles.

## Materials and Methods

Our laboratory test procedure involved preparing a rough fracture, measuring the morphology of halves by a scanner laser, and measuring permeability. The rock used was granite with a unit weight of 25.9 kN/m$^3$ and a uniaxial compressive strength of 172 MPa. An artificial rock joint was created by splitting the specimen mid-height with a special joint-creating apparatus, which has two horizontal jacks and a vertical jack [18-19]. The sides of the joint were cut down after it was created. The final size of the sample was 180 mm in length, 100 mm in width, and 80 mm in height. A virtual mesh with a square element size of 0.2 mm was spread on each surface, and the height at each position was measured with a laser scanner. The procedural details of reconstructing the aperture fields can be found in (Sharifzadeh, 2005; Sharfizadeh at al, 2008) While recording the variation of surfaces, different cases of normal stresses (1, 3, and 5 MPa) were used (Figure 1). Next, the non-contact areas were mapped onto a network. To set up a network, we considered each aperture profile as a node. Each profile has *N* pixels, with each pixel showing the void size of that cell. Depending on the direction of each profile, its length can be observed to change. In our study, the maximum number of profiles were observed in the direction perpendicular to the shear, while the minimum number was found in the parallel direction. To make an edge between two nodes, a correlation measurement over the aperture profiles was used. For each pair of profiles $V_i$ and $V_j$, containing N elements (pixels), the correlation coefficient can be written as:

$$C_{ij} = \frac{\sum_{k=1}^{N}\left[V_i(k) - \prec V_i \succ\right].\left[V_j(k) - \prec V_j \succ\right]}{\sqrt{\sum_{k=1}^{N}\left[V_i(k) - \prec V_i \succ\right]^2}.\sqrt{\sum_{k=1}^{N}\left[V_j(k) - \prec V_j \succ\right]^2}} \quad (1)$$

where $\prec V_i \succ = \frac{\sum_{k=1}^{N} V_i(k)}{N}$. We constructed networks from the measured apertures following two directions: parallel and perpendicular to the shear direction. To make an edge between two nodes, a correlation measurement ($C_{ij}$) over the aperture profiles was used. We assumed that if $C_{ij} \geq r_c$, then a link between two nodes was attached. To choose the optimum value $r_c$ (or a range of that), we note that the aim is to reach or keep the most stable structures in the total topology of the constructed networks. Different approaches have been used to this effect, focusing on the density of links, the dominant correlation among nodes, and distribution of edges or clusters. To choose $r_c$ , we used a nearly stable region in the betweenness centrality (B.C) - $r_c$ space (Fig. 1d), which is in analogy with the minimum value in the rate of edge density (Fig. 1c) (Gao et al, 2009). Betweenness centrality is a measure of how many shortest paths cross through a node. This method has been used successfully in analysis of time-series patterns in network spaces. This interval is nearly equal to a choice of $r_c$ to be about 10-20 % of the maximum correlation value. The mentioned aperture patches may be observed in a manifold space. By combining several patches (profiles) – from several lines/observers like 499 for parallel and 890 for perpendicular – we can reach a reconstructed “phase space” for a certain time step. In other words, we can interpret the overall observations in terms of time-delay coordinates. Following this, we can relate the aperture space analysis to a time series analysis with complex networks. The latter interpretation of the profiles (either aperture or roughness profiles) can be considered a novel way in the analysis and interpretation of fracture surface topography and friction patterns analysis.

To proceed, we used several characteristics of networks. Each node is characterized by its degree $k_i$ and the clustering coefficient. The clustering coefficient as a fraction of triangles (3 point loops/cycles) is $C_i$ defined as $C_i = \frac{2T_i}{k_i(k_i - 1)}$ where $T_i$ is the number of links among the neighbors of node *i*. It follows that a node with *k* links participates on *T(k)* triangles. For a given network with *N* nodes, the degree of the node and Laplacianof the connectivity matrix are defined by (McGraw et al, 2008; Samukhin et al ,2008) :

$$k_i = \sum_{j=1}^{N} A_{ij} ; L_{ij} = A_{ij} - k_i \delta_{ij} \qquad (2)$$

where $k$ , $A_{ij}$ , $L_{ij}$ are the degree of $i$ th node, elements of a symmetric adjacency matrix, and the network Laplacian matrix, respectively. The eigenvalues $\Lambda_\alpha$ are given by $\sum_{j=1}^{N} L_{ij}\phi_j^\alpha = \Lambda_\alpha \phi_i^\alpha$ , in which $\phi_i^\alpha$ is the $i$ th eigenvector of the Laplacian matrix ($\alpha = 1,...,N$). With this definition, all eigenvalues are non-positive values. The inverse participation ratio as a criterion of the localization of eigenvectors is defined by (McGraw et al, 2008):

$$P(\varphi^{\alpha}) = \frac{\sum_i \phi_i^{\alpha}}{(\sum_i (\phi_i^{\alpha})^2)^2}; \alpha = 1, ..., N \quad (3)$$

The maximum value of ***P*** shows that the vector has only one non-zero component. A higher value of *P* corresponds with a more localized vector. To understand the nature of energy propagation in obtained friction-networks, we activated all the nodes simultaneously (*i.e.*, a Kuramato model in connected oscillators (Kuramoto, 1984). Diffusion of information ( $u_i$ ) is expressed by a network mode of diffusion (Nakao and Mikhailov, 2010; Arenas et al, 2008):

$$\frac{d}{dt} u_i(t) = f(u_i) + \varepsilon \sum_{j=1}^{N} L_{ij} u_j \quad (4)$$

where $f(u_i)$ and $\varepsilon$ are local dynamics of each profile (node) and a diffusion constant, respectively (in our simulation $f(u_i) = 0; \varepsilon = 1$). We used the synchronization criterion of randomly (uniform) stimulated of nodes by $\mathrm{M} = \lim_{t \to \infty} \langle | u_i(t) - u_j(t) | \rangle_{i,j}$ in which <...> denotes an average value over all nodes after passing significant time steps. Here, we assumed a frictional interface, which is pushed under a constant driving stress. The storied energy in asperities is proportional with elastic properties of asperities and contact areas. The changes in contacts are expressed in terms of friction laws such as rate and state or Coulomb's friction laws (Dieterich, 1978; Ruina, 1983). If we consider that there is a long-range correlation among particles (elements) of the interface, then the patterns of energy propagation (and storage) will be dramatically affected by correlation patterns. To better understand the nature of energy propagation, we note the importance of local structures, determined by sub-graphs. Analysis of the internal network structures is presented by sub-graphs and motifs. The sub-graphs are the nodes within the network with the special shape(s) of connectivity together. The relative abundance of sub-graphs has been shown to be an index to the functionality of networks with respect to information processing. Also, they correlate with the global characteristics of the networks (Vazquez et al, 2004; Boccaletti et al,2006; Xu et al,2008). The network motifs introduced by Milo et al (Milo et all, 2002 and 2004) are particular sub-graphs representing patterns of local interconnections between the nodes in the network. A motif is a sub-graph that appears more than a certain amount (further criteria can be found in secondary literature). A motif of size *k* (containing k nodes) is called a *k*-motif (or generally sub-graph). We employ the aforementioned approaches over the networks of aperture profiles.

## Results and Discussion

In Figure 2, we plotted the three parameters of the networks through 20 mm shear displacements. After a transition stage, the formation of 3-node loops (clustering coefficient) in a parallel direction reaches a quasi-stable state, while the perpendicular networks follow a growing trend (Fig. 2a). This trend is reversed in the evolution of the nodes' degree (Fig. 2b), which shows the growth of the profiles' long-range correlations. The characteristic length of networks exhibits a rapid drop after passing the interlocking step, where the asperities are locked up. This stage occurs just after the peak point of shear stress-displacement. However, for the perpendicular-direction case analyzed, the transition point at the same stage displayed a softer change (Fig. 2c). The similarities among the three presented characteristics of the networks occur in the transformation from 1 mm to 2 mm of displacement, where the rock joint under a certain value of normal stress passes the maximum frictional strength (see Figure 1; case 3). Furthermore, a consideration of 3-point cycles (*T*-triangle loops) versus the nodes' degree shows a power low scaling (Fig. 2g, h):

$$T(k) \sim k^{\beta}, \tag{5}$$

where the best fit for the collapsed data set reads $\beta \approx 2 \pm .3$ (which we call a coupling coefficient of local and global structures). With some mathematical analysis (see appendix), one can show that adding *m* edges increases the number of loops with $\beta^2 m^{\beta}$, which indicates a very congested structure of global and local sub-graphs during shear rupture. Also, we notice $C(k) \sim 2k^{\beta-2}$, so that for $\beta < 2$, a possible hierarchical structure can be predicted (Albert and Barabási ,2002) . Let us consider a simplified form of the standard equation for the friction based on the rate and state friction law (Dieterich, 1978) , with an assumption of nearly constant sliding velocity: $\tau \sim \sigma \ln \frac{\theta}{D_c}$, in which $\theta$ is the variable describing the interface state, and $D_c$ is the characteristic length for the evolution of $\theta$ ($\tau$ is shear stress and $\sigma$ is normal stress on the interface). The commonly used empirical laws for evolution of state variable are Ruina's laws for ageing and slipping states (Ruina, 1983) . For slip law it reads (with assuming $v \equiv 1$): $\frac{\partial \theta}{\partial t} \sim -\frac{\theta}{D_c} \ln \frac{\theta}{D_c}$ . Let us transfer the state variable in terms of local and global characteristics of the interface:

$$\frac{\partial \theta_i(t)}{\partial t} = a\frac{\partial k_i}{\partial t} + b\frac{\partial T_i}{\partial t} \qquad (6)$$

which we assumed that the evolution of the state variable is associated with the evolution of local and global parameters of friction network. We eventually obtain (with plugging 5 in 6):

$$\frac{\partial \theta_i(t)}{\partial t} = \frac{\partial k_i}{\partial t}(a + b\beta k_i^{\beta-1}) \qquad (7)$$

This relation indicates that with the assumption of an evolution law for the obtained networks, one can interpret friction laws in terms of local and global structural complexities. For example, let us assume that our network includes some “hub nodes” that tend to absorb more loops; in other words aggregation of loops around hubs. Assuming preferentiality attachment (or detachment (Boccaletti et al ,2006; Albert and Barabási ,2002 ), such a case leads to:

$$\frac{\partial T_i(k)}{\partial t} \sim m\frac{T_i}{\sum T_j}, \qquad (8)$$

in which *m* is a coefficient of growth (or decay). Plugging (5) in (8) yields:

$$\frac{\partial k_i}{\partial t} \sim \frac{m}{\beta}\frac{k_i}{\sum k_j^{\beta}} \qquad (9)$$

For $\beta = 1$, the model yields scale-free networks (Albert and Barabási, 2002) . Assuming $\beta = 1$ and $a < 0$ indicates a decaying model for the state parameter in terms of attacking to hubs. Plugging (9) into (7) and assuming $\beta \approx 2$ leads to:

$$\frac{\partial \theta_i(t)}{\partial t} = \frac{2bmk_i^2}{\sum k_j^2} + \frac{amk_i}{\sum k_j^2} \qquad (10)$$

which shows a complex non-linear decaying nature of state parameters. The first term in right hand of Eq.10 is a non-linear evolution of network with gel-like characteristic, i.e. a single node is connected to almost all nodes in the friction network. Further development of (10) with respect to different network models will be addressed elsewhere.

In Figure 3, we compared the results of experimental measurements of hydraulic conductivity with the inverse of the characteristic length in parallel profile networks (Figures 3a, b). The same temporal evolutionary trend can be observed between the inverse of the parallel aperture profiles' mean geodesic length and the measured hydraulic conductivity. As we have shown in Figure 4, the evolution of mean geodesic length coincides with the formation of clusters over the parallel networks. We have found that the rate of propagation of information is much higher in the last evolutionary steps, as compared to the initial development of the rock joint. This difference is due to the congestion of contact areas, which induce the trapping of the energy (Bowden and Tabor, 2001; Rubinstein et al, 2004).

Comparing the number of edges in the parallel direction in a semi-logarithmic graph shows the same trend for the evolution of measured hydraulic conductivity (Figure 3c). From Figure 4, it is clear that after a transition step the concentration of edge growth is on certain profiles. This finding seems to indicate the concentration of energy flow which is comparable with betweenness centrality. This phenomenon is related to the paths of fluid flow with the fracture and through porous spaces (channelization). Observation of the same trend in the inverse of the mean characteristic path and the permeability evolution distinguishes the formed dynamic groups' roles over the obtained networks. To show the revealed groups' evolution in friction networks, we use joint degree distribution for both cases (parallel and perpendicular profiles). This shows assortativity or hubness characteristics in the networks' topology. In Figures 5 and 6, we show the frequency of the joint degree distribution (the probability of finding an edge with having two specified values of nodes degree) for parallel and perpendicular profiles through the displacements. The main appearance of the distinguished groups occurs shortly after the slip-weakening distance (in this case around 5mm). Generally, the evolution of groups for both cases is almost identical. This means that in the last stages, the nodes with high and low degree connect to the high and low degree nodes, respectively. Now, we analyse eigenvalues and eigenvectors of a Laplacian matrix of networks, where our primary interests are the localization of eigenvectors and the patterns of eigenvalues spectrum.

Evaluation of the localization of eigenvectors for perpendicular networks (based on Eq. 3 – Figure 7) shows that localization is propagating toward the inside of profiles from their boundaries, while in parallel networks, after the interlocking of asperities (SD~1mm) localization is following nearly random patterns. For both cases, the boundary profiles are much more localized than interior profiles. The localization of eigenvectors is more remarkable in parallel profiles than perpendicular ones. This confirms the Heisenberg localization principle, which states that localization in data (here the development of a frictional interface) can be related to localization processes in a kernel matrix (adjacency matrix) spectrum (Coifman et al, 2005; Nakao and Mikhailov, 2010) . Implementation of Eq. 4 in the perpendicular friction networks proved that the synchronization of information in initial stages of the evolution is nearly 1000 times slower than it is in the final, quasi-stable stages (Figure 8). It is noteworthy that the trends of *M* with shear displacements (SD) are the same as the temporal evolution of the characteristic length. In other words, the characteristic length of friction networks can be used as an index to the rate of radiated storied energy. Shorter characteristic length indicates faster radiation of energy. Perpendicular networks are characterized by the *M* value's fast synchronization and long range of variations, while the distance between *M* values in pre- and post-peak stages is nearly one order. In our case study, it follows that information propagation (entrapping of energy) through profiles is faster in perpendicular profiles than in parallel ones; the rate of energy storage is controlled with perpendicular patterns of contact patches. Here, one may assume that the flux of energy is storied over the networks, while the topology of the networks is invariant during the entrapping or deformation process Depending on the network structure, the synchronization time (time until a steady state is reached) will be different (Arenas et al ,2008; Oh et al,2005). One may infer that the trapping of energy in pre-peak stages is faster than in post-peak stages. This is completely reasonable in terms of the complex configuration of the contact areas where it occurs before slip-weakening stage (unstable stage). In parallel networks, the difference between maximum and minimum ranges is within one order, while for perpendicular networks, it is nearly 3 orders. This can be another reason to the importance of perpendicular nature of friction networks and their answer to environment stimuli.

Figure 9 shows the analysis of 4-point sub-graphs over both perpendicular and parallel networks. An increase of the abundance of sub-graph patterns in SD=2mm to 3mm (Figure 9a) is in agreement with the mechanical deformation and dilatancy of the fracture. Also, the drop in all sub-graphs from SD=0mm to SD=1mm (Figure 9 b) is the index of an interlocking and dramatic drop of the permeability. The common property for both perpendicular and parallel sub-graphs is the main evolutionary trend of sub-graphs. For example, after slip, the transient sub-graphs (index 6) show faster growth and percolation over both networks. The same evolutionary trend is predicted for directional profiles (neither parallel nor perpendicular), indicating an identical mechanism in the functionality of networks (the same mechanism in fracture evolution). The rapid increment of index 6, notably for parallel networks, is in absolute agreement with the easy fluid flow in the residual stages of the sheared fracture. The low value of index 2 and 4 shows the localization of flow, i.e., the channelization effect. The appearance of index 2 and 4 is much more relevant to flow heterogeneity. It follows that most transient sub-graphs resembling index 6 are accompanied by much more stable flow patterns. We tested our approach on three other cases, where normal stress values were varied. For all cases, the results of the 4-point sub-graph analysis showed that index 4 and, relatively speaking, index 2 displayed the minimum frequency (Ghaffari et al, 2011b). Accordingly, we observed a similar trend, a super-family phenomenon, for a variety of different cases of shear rupture in sub-graphs distribution.

## Conclusion

In this study we characterized the spatial structural complexity of apertures using networks and the idea of the long-range correlation of patterns. We analysed friction-networks and tried to link the complex friction patterns to networks parameters. This led to a linking between classical friction formulations and the corresponding network parameters. The characteristics of the networks scaled with mechanical and hydraulic properties of a frictional interface. Information propagation through profiles was found to be much faster in perpendicular profiles, as compared to parallel profiles. We found that there is a universal power law between the degrees of nodes and sub-graph frequency, which indicates a fast localization of clusters around hubs. Based on this scaling law and state-rate friction relations, we developed a mathematical framework for friction-networks. Comparison of synchronization patterns with the localization of eigenvectors from the Laplacian of the connectivity matrix showed an inverse relation between localization and synchronization. From another point view, it seems that localized eigenvectors are leading most of the energy flux. We found the same temporal evolutionary trend in the inverse of the mean geodesic length of the parallel aperture profiles with the measured hydraulic conductivity (upon different inlet water pressures). The latter observation confirmed the role of huge clusters (groups) over the parallel networks in conducting information. Motif analysis of different cases in shear rupture confirmed the same inherent dynamic of sheared fracture, which yields a nearly identical family of sub-graphs. The trend of different sub-graphs roughly correlated with the fluid flow features. More analysis is needed to better understand sub-graph distribution and other properties of fluid flow in sheared fractures.

❖ We would like to acknowledge and thank Prof. M. Sharifzadeh (Tehran Polytechnic) and Prof. E. Evgin (Unversity of Ottawa), who shared their comments and suggestions during the preparation of the manuscript. Special thanks to M. Sharifzadeh for providing the data set employed in this study. We thank the editor J. Kurths and two anonymous referees for useful comments and suggestions on the paper.

## Appendix

We show that adding a link to obtained networks increases the number of 3-point cycles with a factor of $\beta^2$.

$$T(k) \sim k^{\beta}$$

$$T(k+1) \sim (k+1)^{\beta} = k^{\beta}(1+\frac{1}{k})^{\beta} \equiv \overbrace{k^{\beta}}^{\sim T(k)} \sum_{n=0}^{\infty} \binom{\beta}{n} k^{-n}$$

$$\frac{T(k+1)}{T(k)} \sim \sum_{n=0}^{\infty} \binom{\beta}{n} k^{-n} \approx \sum_{n=0}^{\infty} \frac{\beta(\beta-1)...(\beta-n+1)}{n!} k^{-n} \approx \beta^2(1+\frac{\beta}{k})$$

For large $k$, It reads : $\frac{T(k+1)}{T(k)} \sim \beta^2$.

# Figures

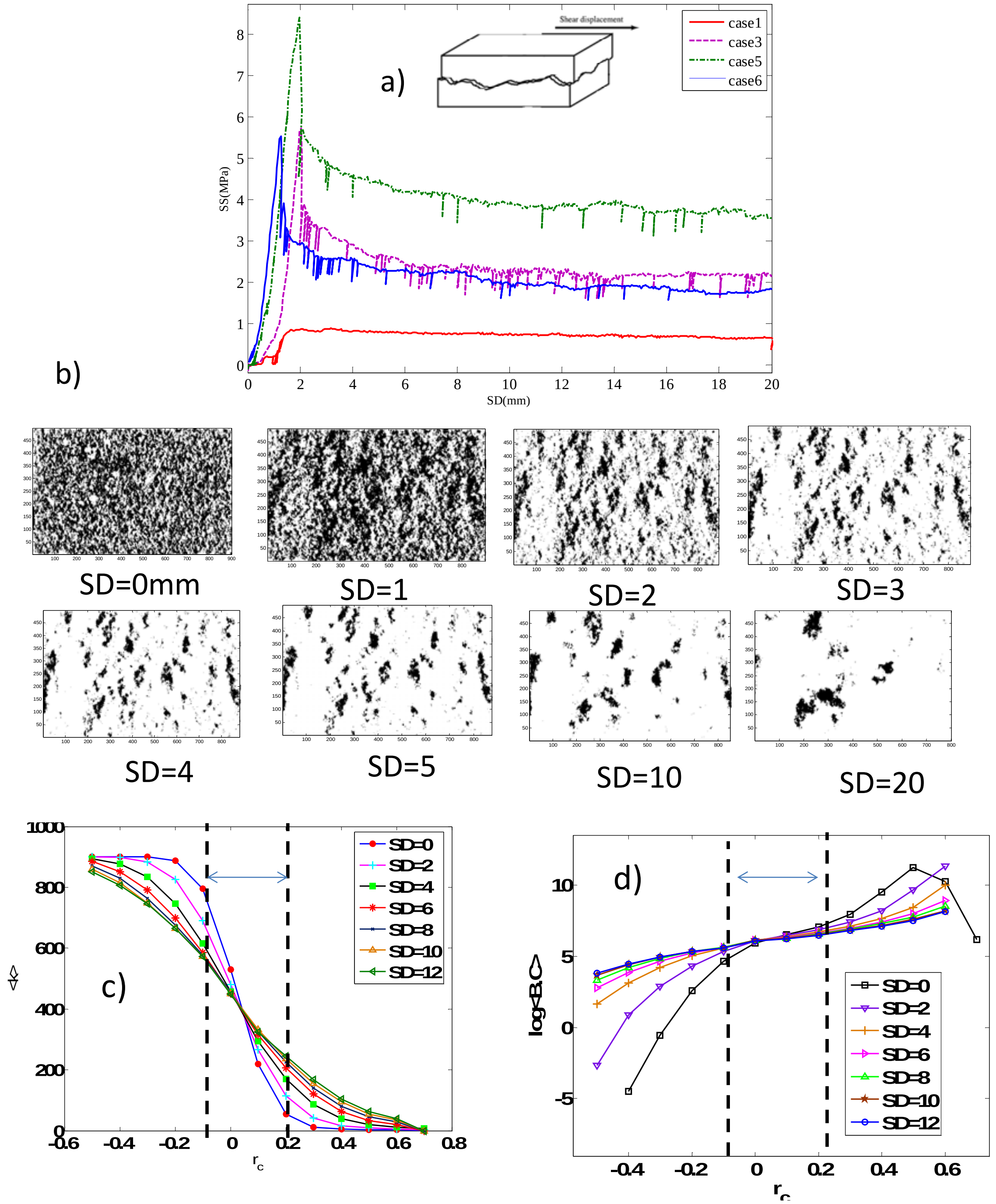

Figure 1. a) Shear strength for different cases (SD=Shear Displacement in mm; normal stresses for case1:1Mpa, case3: 3 MPa, case5: 5 MPa and case6: 3 MPa (without control of upper shear box); b) Contact zones (black) and non-contact areas during the evolution of a frictional interface (case 3) ;(c) density of edges versus threshold level and d) natural logarithmic variation of mean betweennness centrality (B.C) with truncation value. The indicated interval with arrows shows the best possible threshold level where the minimum variation of log<B.C> occurs (the most stable-dominant structures). <...> indicates average over all nodes (i.e., aperture patches).

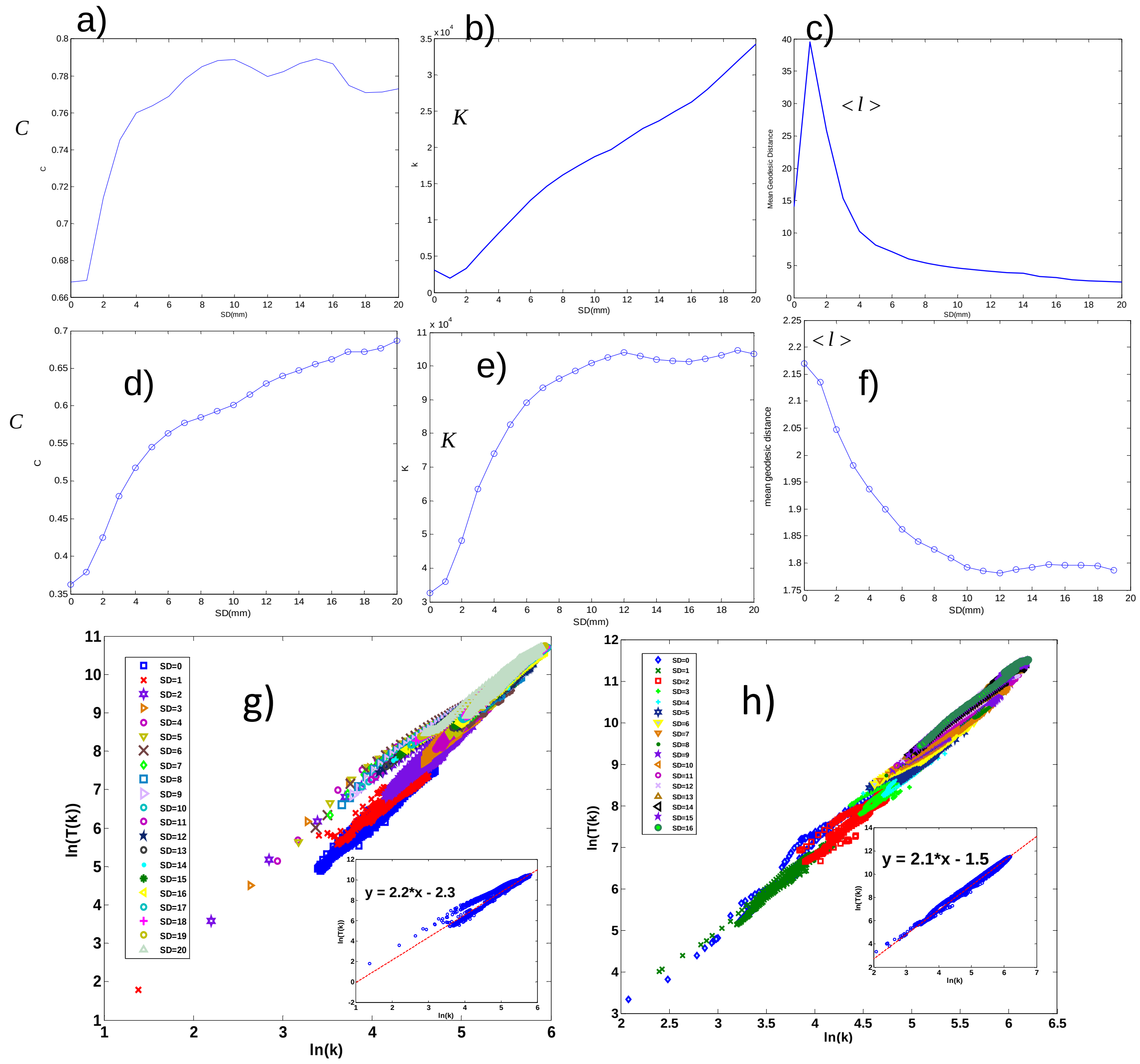

Figure 2. Characteristics of the friction-networks in parallel profiles-networks to shear direction (a -c) :a) Clustering coefficient-Shear Displacement (SD in *mm*);b) Number of edges-SD ; c) Average path length-SD; in transverse profiles-networks to shear direction (d -f) ; (g) X-profiles :scaling of triangles (*T(k)*-i.e., loops) with node's degree(*k*) as a power law with $\beta \approx 2.2$ (inset shows the best fit linear line to collapsed Data set in natural logarithmic scale) .Also, We confirmed for real-time contact measures there is such universal power law [33];(h) Y-profiles (parallel to shear) a nearly same scaling with $\beta \approx 2.1$.

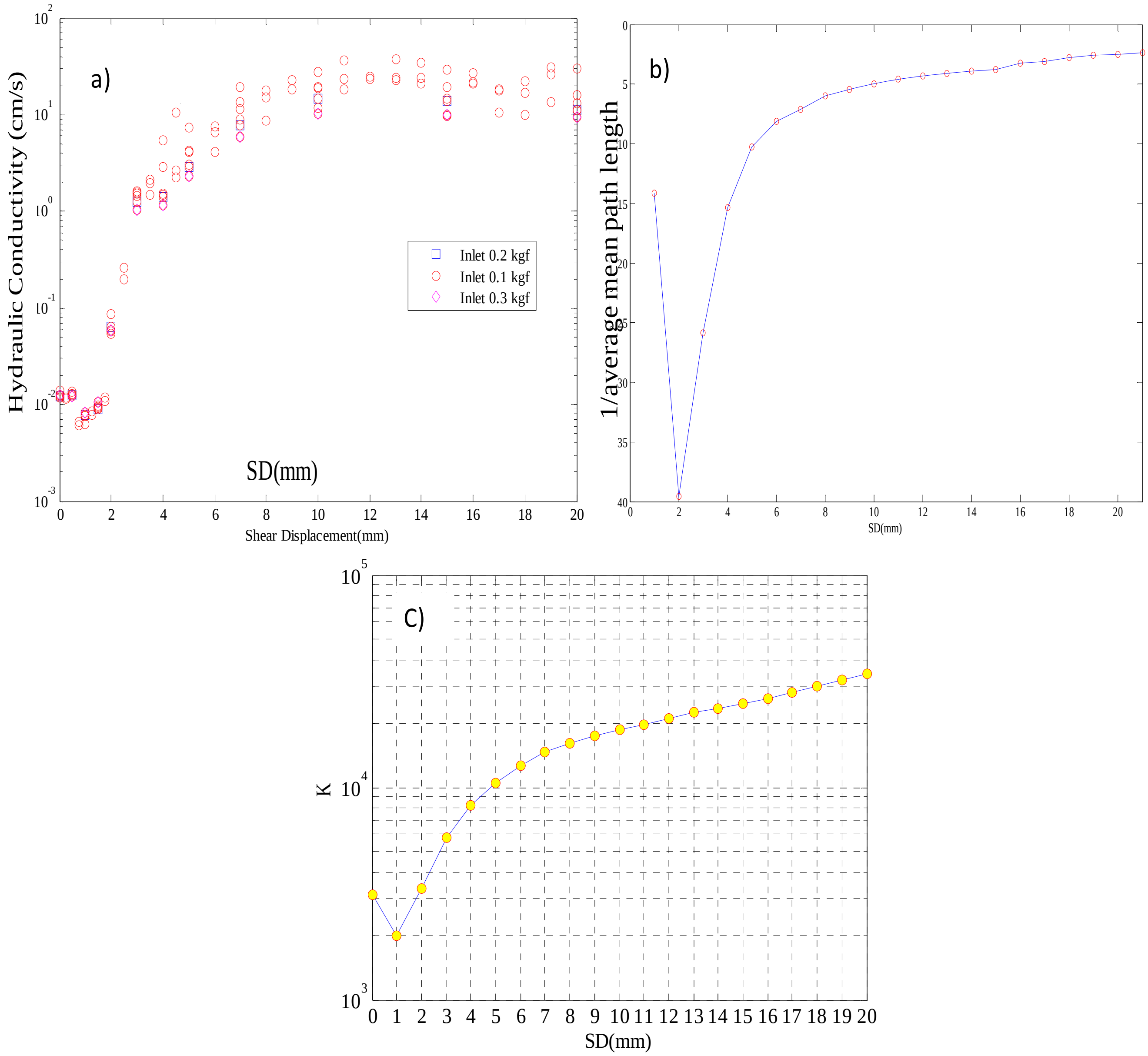

Figure 3. Comparison of (a) hydraulic conductivity (measured through experimental laboratory tests) and (b) the inverse of the mean geodesic length of the parallel aperture networks (flipping Y-axis) and (c) evolution of edges degree with displacements in semi-logarithmic scale.

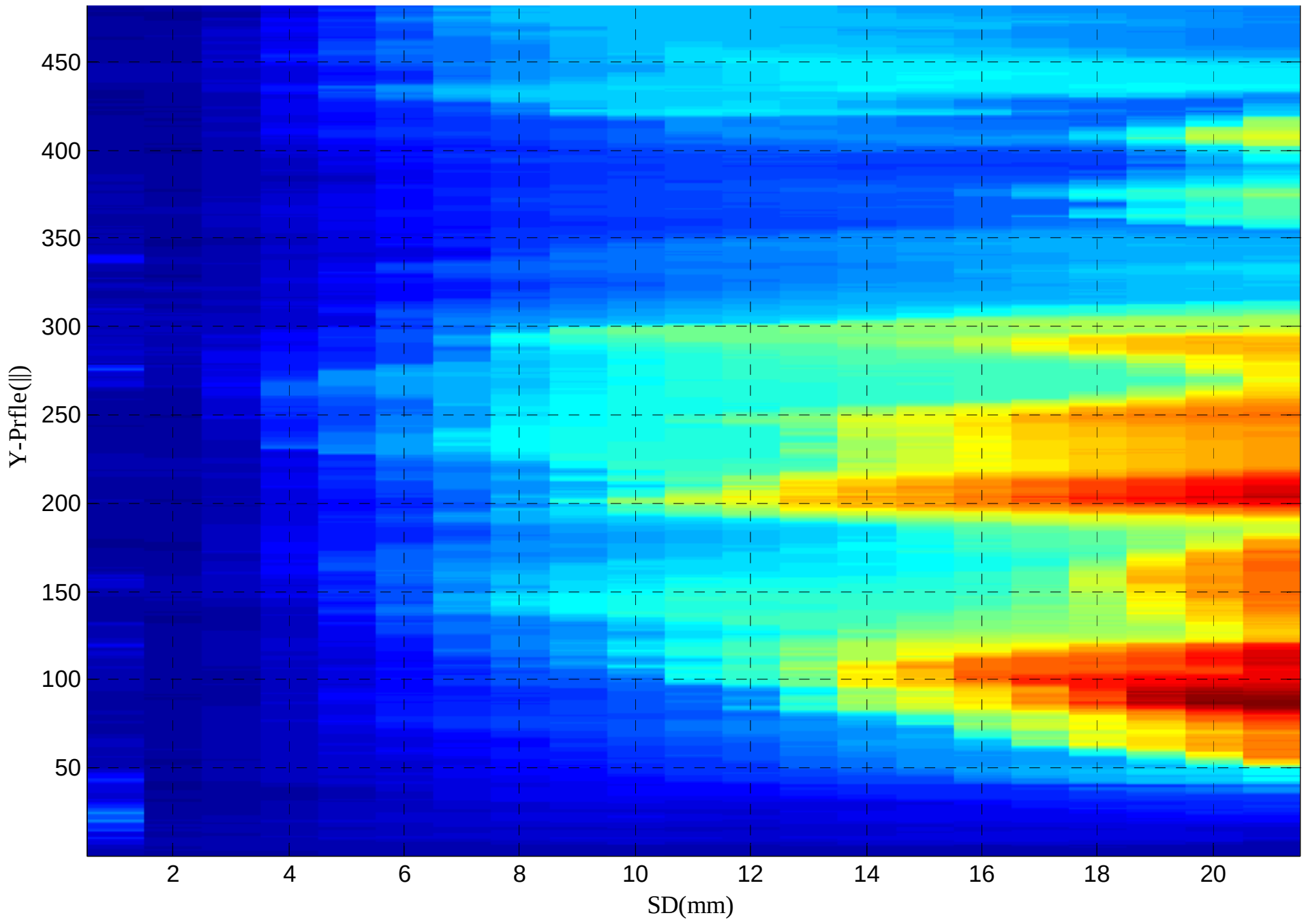

Figure 4. Development of the logarithmic scale of the node’s degree over parallel-aperture profiles (vertical axis is proportional with Y- axis) with shear slips (SD: in mm - horizontal axis). Rapid growth of K= $\frac{1}{2}\sum_i k_i$ occurs in portion of aperture patches.

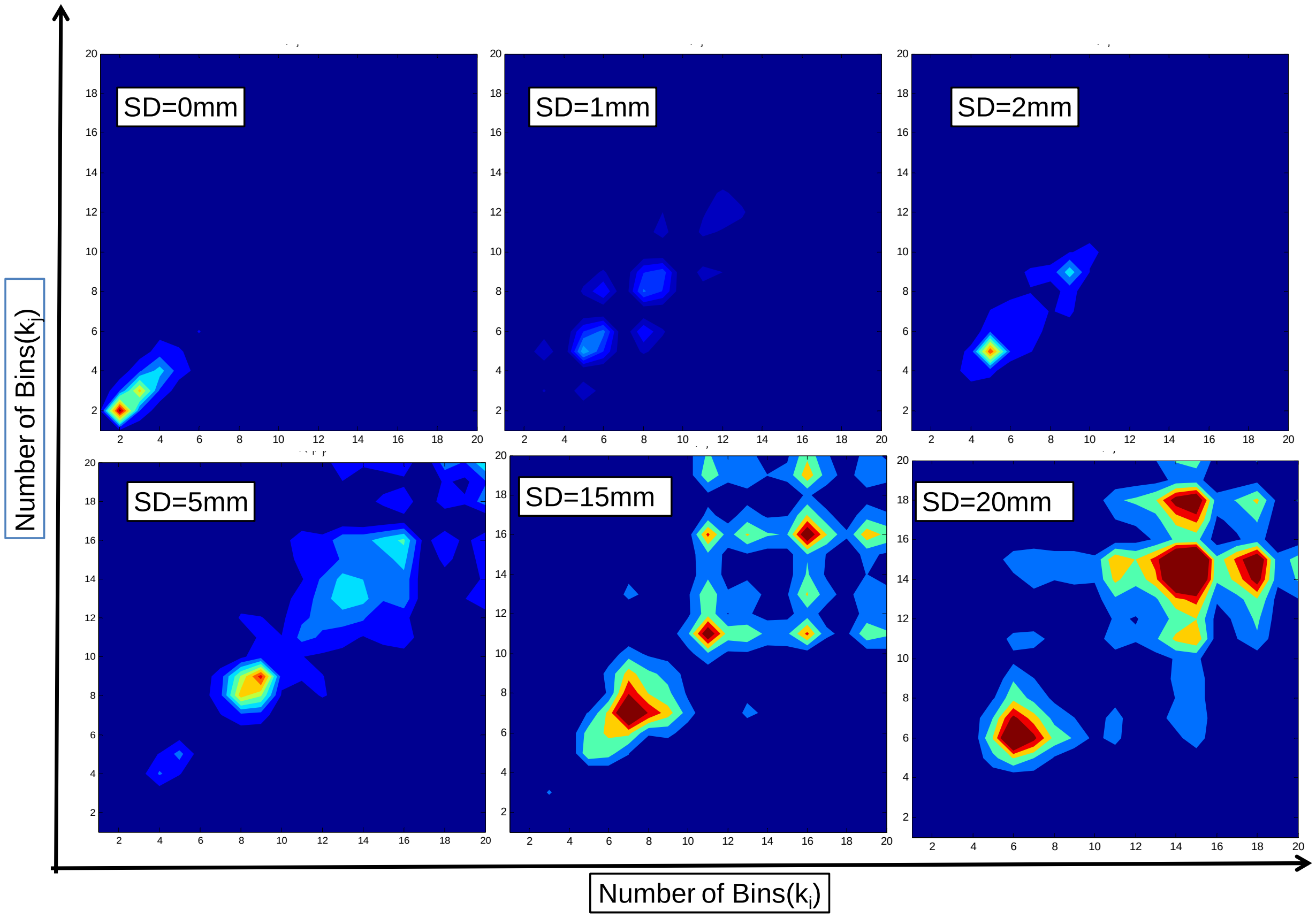

Figure 5. Assortative networks in parallel-friction networks: Joint degree distribution evolution through the successive displacements of the rock joint (parallel profiles) from shear displacement 0 mm to 20 mm.

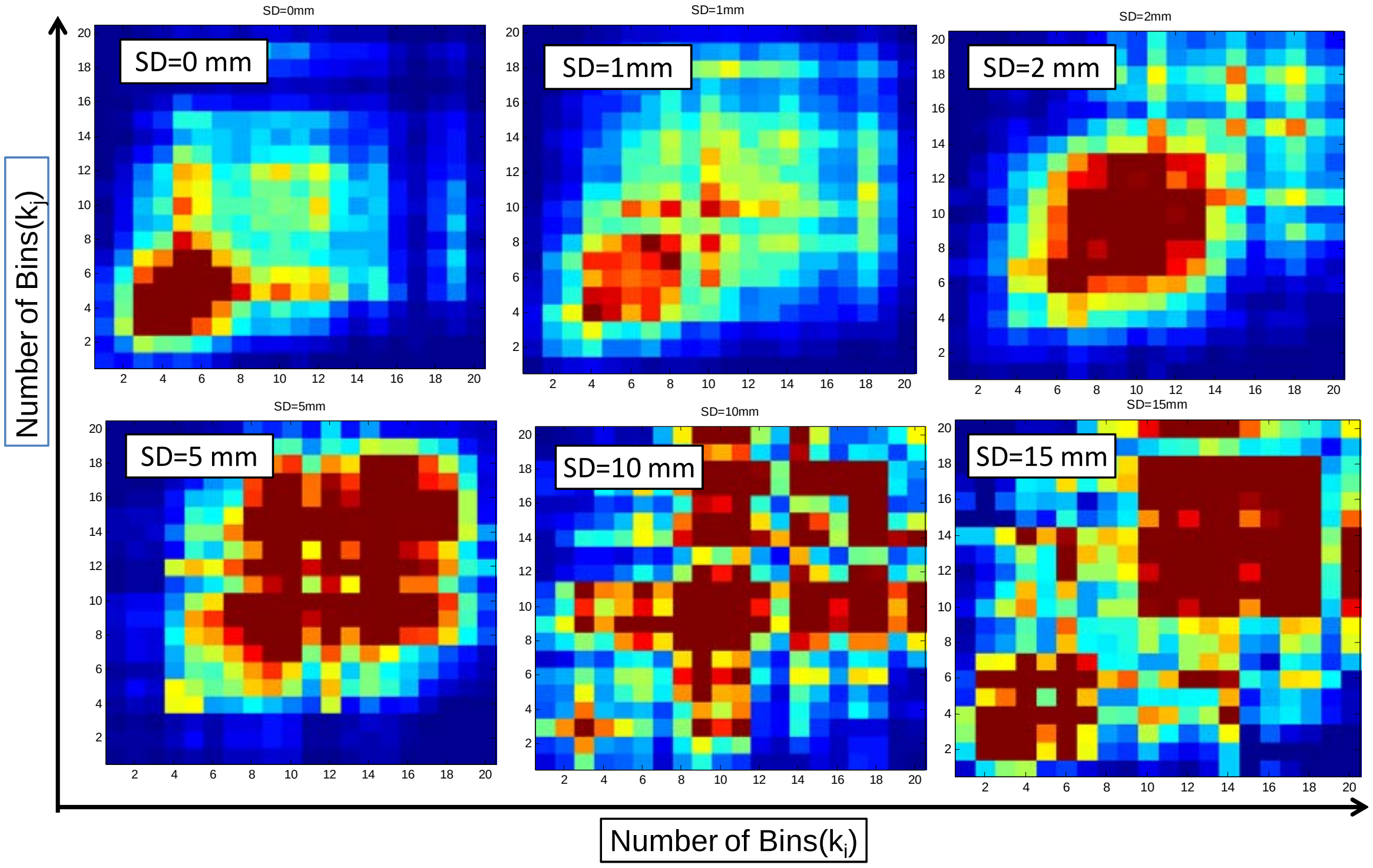

Figure 6. Joint degree distribution evolution for perpendicular aperture-friction networks from SD=0mm to SD=15 mm.

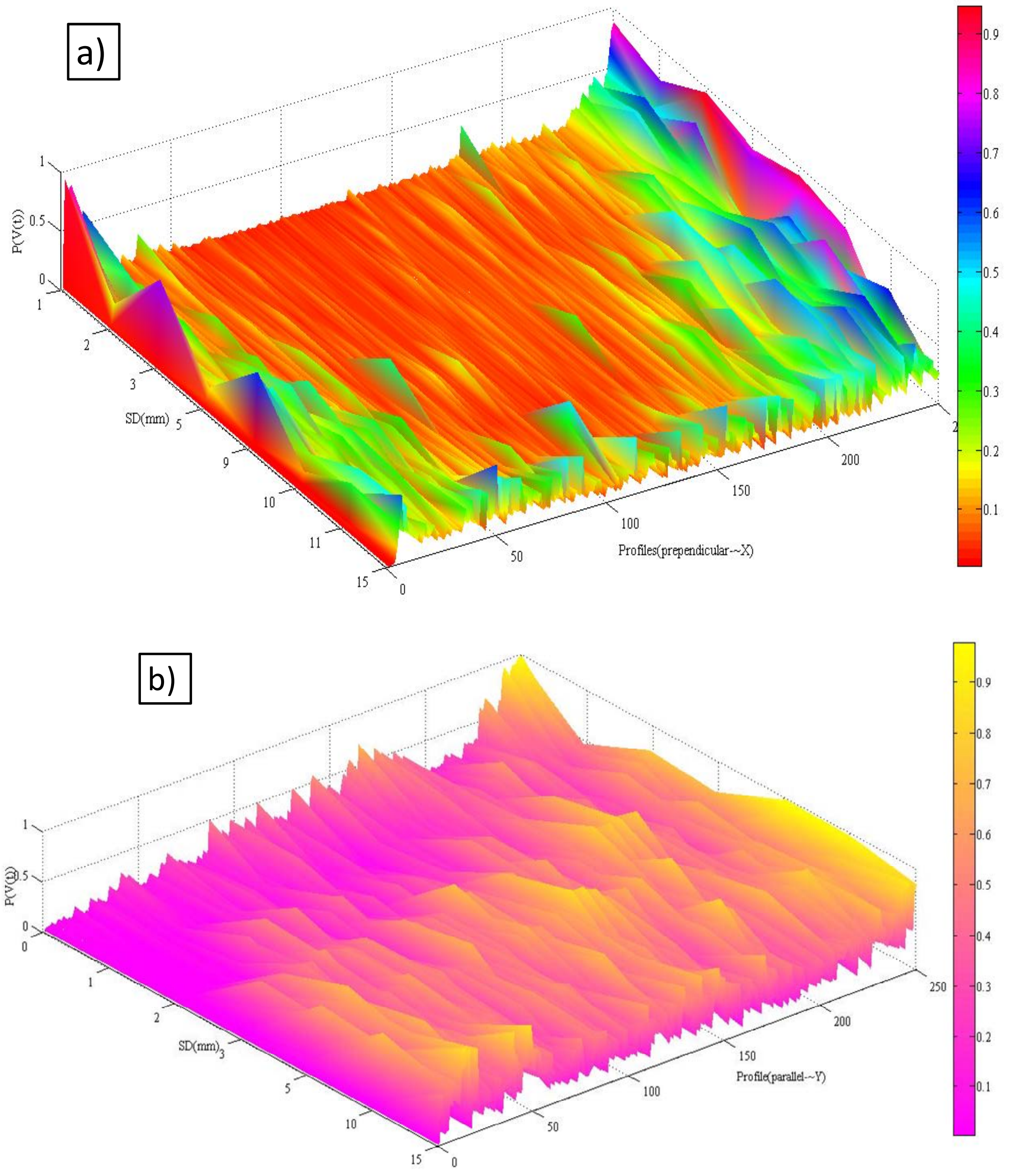

Figure 7. Evolution of the inverse participation factor for profiles in X and Y directions :a) perpendicular profiles are localized from boundaries to inside profiles. b) Parallel profiles after interlocking of asperities (SD~1mm) are nearly following random localization.

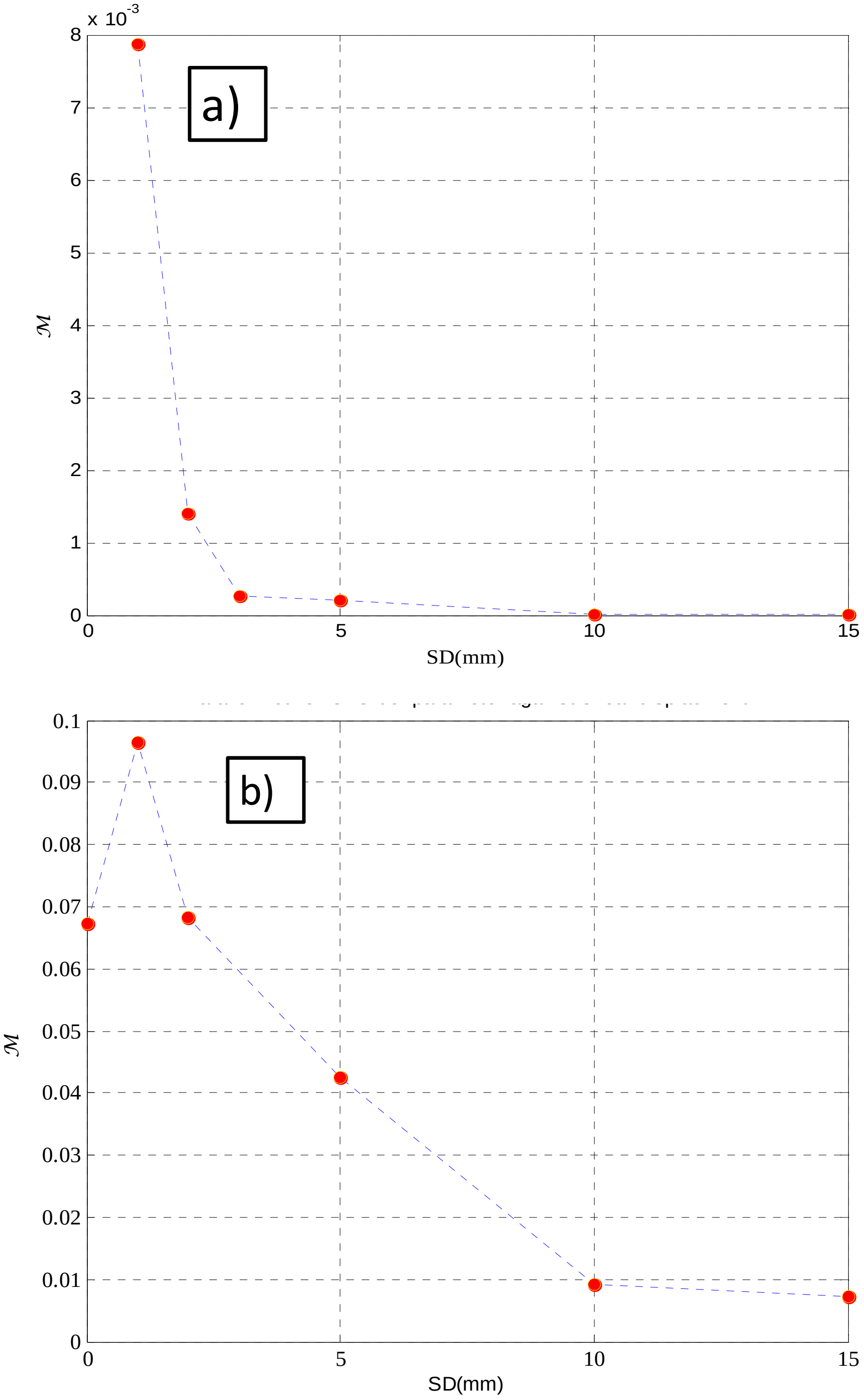

Figure 8. Convergence of the synchronization criterion (M) over 100 realizations (for each Shear Displacement –SD (mm)) of diffusion based networks equation on a) perpendicular and b) parallel aperture networks. The number of nodes in the model for all of the cases is decreases to 250 nodes (1:4 scaling).

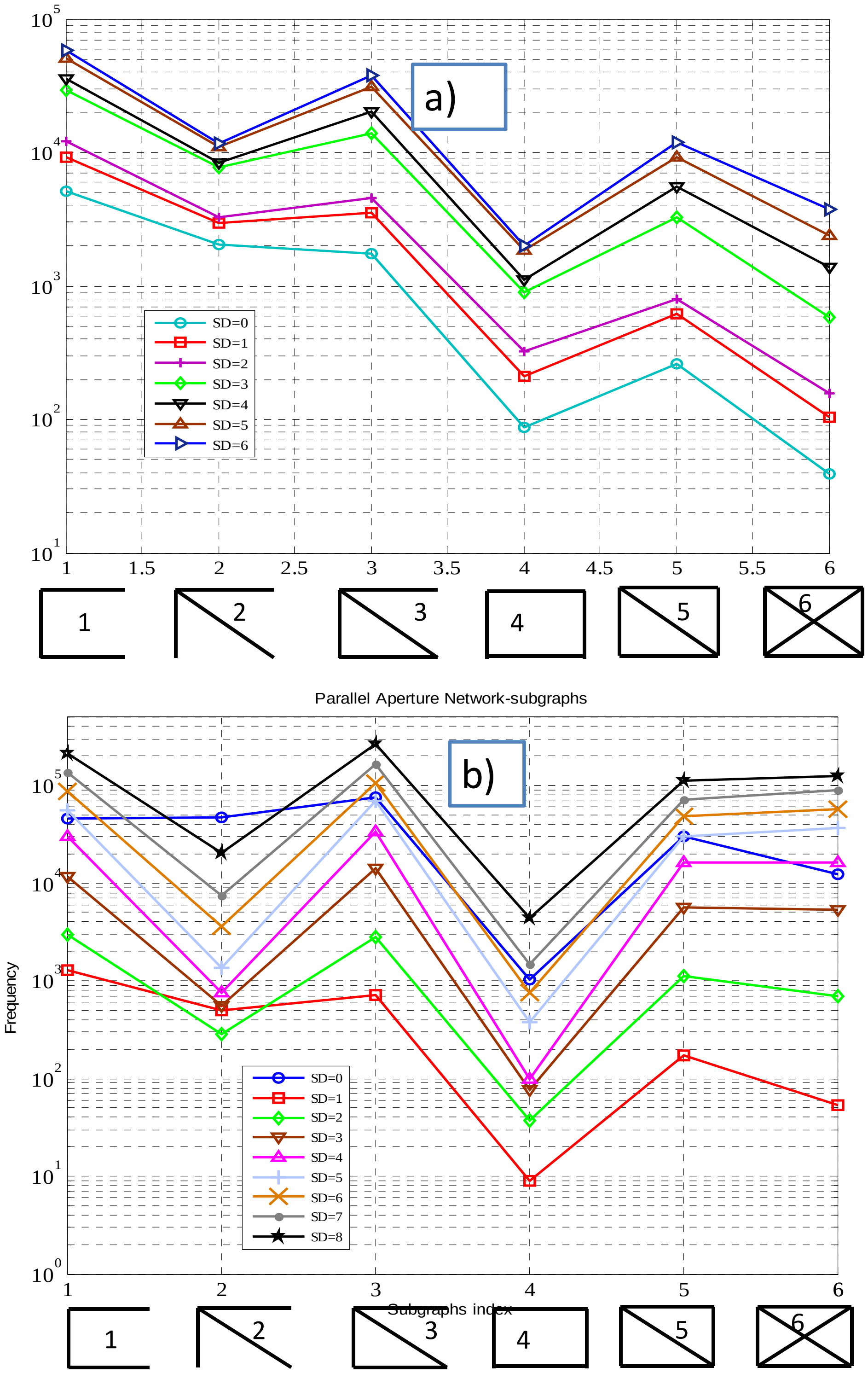

Figure 9. Evolution of 4-points sub-graphs for profiles in X and Y directions though each shear slip (SD, mm): a) perpendicular profiles, b) Parallel profiles.